\documentclass[11pt]{article}
\usepackage{moriond,epsfig,graphics}

\def\be{\begin{equation}}
\def\ee{\end{equation}}
\def\bea{\begin{eqnarray}}
\def\eea{\end{eqnarray}}

\newcommand\aj{AJ}%                                         % Astronomical Journal
\newcommand\mnras{MNRAS}%                                   % Monthly Notices of the RAS

\begin{document}
\vspace*{4cm}
\title{EFFECT OF DISTANCE ERRORS: APPLICATIONS TO SDSS EARLY-TYPE GALAXIES}

\author{GRAZIANO ROSSI and RAVI K. SHETH}

\address{Department of Physics and Astronomy, University of Pennsylvania, PA 19104, USA}

%%%%%%%%%%%%%%%%%%%%%%%%%%%%%%%%%%%%%%%%%%%%%%%%%%%%%%%%%%%%%%%%%%%%%%%%%%%%%%%%%%%%%%%%%%%%%%%%%%%%%%%%%%%%%%%%%%%%%%%%%%

\maketitle
\abstracts{
Noisy distance estimates associated with photometric rather than
spectroscopic redshifts lead to a mis-estimate of the
luminosities, and produce a correlated mis-estimate of the sizes.
We consider a sample of early-type galaxies from the SDSS DR6 and apply the
generalization of the $V_{max}$ method to correct for
these biases. We show that our
technique recovers the true redshift, magnitude and size
distributions, as well as the true size-luminosity relation. 
Regardless the specific application outlined here, our method
impacts a broader range of studies, when at least
one distance-dependent quantity is involved.}

%%%%%%%%%%%%%%%%%%%%%%%%%%%%%%%%%%%%%%%%%%%%%%%%%%%%%%%%%%%%%%%%%%%%%%%%%%%%%%%%%%%%%%%%%%%%%%%%%%%%%%%%%%%%%%%%%%%%%%%%%%

\section{Introduction and significance}
Galaxy scaling relations play a
crucial rule in constraining galaxy formation models.
However, a bias will be intrinsically present in these correlations
if the transformation from observable to physical quantity involves
one or more distance-dependent observables, due to noise in the
distance estimate. Distances are only known approximately if photometric redshifts are
available but spectroscopic redshifts are not.
This is already the case of many current surveys (e.g. SDSS, Combo-17,
MUSYC, Cosmos), where the number of objects with photometric redshifts is more than an order of magnitude bigger than
that of spectroscopic redshifts, and will be increasingly true of the next generations of deep
multicolor photometric surveys (e.g. DES, LSST, SNAP).

Therefore, methods for recovering unbiased estimates of
distance-dependent observables, and of the joint
distribution of luminosity, color, size, from magnitude limited photometric redshift
datasets are indeed necessary (Rossi \&
Sheth 2007 \cite{RS2007}; Sheth 2007 \cite{S2007}; Lima et al. 2008
\cite{LCO2007}). 
In what follows, we show the essential inversion character of this class of
problems by using a selected sample of early-type galaxies from the SDSS DR6 -- for which both photo-zs and
spectro-zs are known -- and by applying our deconvolution
techniques to reconstruct the true distributions and the scaling relations. 

%%%%%%%%%%%%%%%%%%%%%%%%%%%%%%%%%%%%%%%%%%%%%%%%%%%%%%%%%%%%%%%%%%%%%%%%%%%%%%%%%%%%%%%%%%%%%%%%%%%%%%%%%%%%%%%%%%%%%%%%%%

\section{The SDSS early-type sample}
The catalog we use is based on the Sloan Digital Sky Survey (SDSS)
Data Release 6, available online through the
Catalog Archive Server Jobs System (CasJobs).
We adopt selection criteria suitable to early-type galaxies.
Specifically, from the DR6 galaxy photometric sample (PhotoObjAll in
the Galaxy view), and from the spectroscopic sample (SpecObjAll), we
select objects according to these general criteria:
\begin{itemize}
\item Petrosian magnitudes in the range $14.50 \le m \le
      17.45$ for the r band;
\item Concentration index
      $r_{petro,90}/r_{petro,50} > 2.5$ in the $i$ band;
\item Likelihood of the de Vaucouleur's model $> 0.8$;
\item Objects with both photometric and spectroscopic redshifts available.
\end{itemize}
No redshift or velocity dispersion cuts were made.
Our catalog contains $163,718$ objects, and consists of model
magnitudes, petrosian radii, De Vaucouleurs and
exponential fit scale radii along with their corresponding axis ratios in the
$r$ band, photometric redshifts and photo-z errors.
We do not apply any K-corrections to our de-reddened model magnitudes, since
our main goal is to test the deconvolution technique
rather than characterize the exact relations.
We select photometric redshifts from the SDSS $Photoz$
Table.
This set of photometric redshifts has been obtained with the template
fitting method, which simply compares the
expected colors of a galaxy with those observed for an individual
galaxy (Budav{\'a}ri et al. 2000 \cite{BSC2000}). 
The spectroscopic pipeline assigns instead a final redshift to each object
spectrum by choosing the emission or cross-correlation redshift with
the highest CL. In the selection of our sample, we tried to minimize
the use of spectral information,
but more robust constraints can be applied in order to reduce
errors in galaxy classification.

%%%%%%%%%%%%%%%%%%%%%%%%%%%%%%%%%%%%%%%%%%%%%%%%%%%%%%%%%%%%%%%%%%%%%%%%%%%%%%%%%%%%%%%%%%%%%%%%%%%%%%%%%%%%%%%%%%%%%%%%%%

\section{Essence of the deconvolution problem}
If we indicate with $\zeta$ and $z$ the photometric and spectroscopic redshifts, respectively,
the problem of estimating the intrinsic redshift distribution $N(z)$
-- normalized number of objects which lie at redshift $z$ -- 
is best thought of as a deconvolution
problem, and if $p(\zeta|z)$ is the probability of estimating the
redshift as $\zeta$ when the true value is $z$, then the distribution
of estimated redshifts is:
\begin{equation}
{\cal N}(\zeta) =  \int N(z)~p(\zeta|z)~{\rm d}z.
\label{pz}
\end{equation}
Equation (\ref{pz}) is an integral equation of the first kind of
the \textit{Fredholm type}, with the conditional probability
$p(\zeta|z)$ as kernel.
A simple iterative scheme proposed by Lucy (1974) \cite{L1974} allows one to reconstruct the
intrinsic distribution after a few iterations,
provided a suitable first guess. 

Similarily, let $M$ denote the true absolute magnitude and $\mathcal M$
that estimated using $\zeta$ rather than $z$. Use $D_{\rm L}(z)$ to
denote the luminosity distance, and $\phi(M)$ to indicate the number density of
galaxies with absolute magnitudes $M$.
Let $V_{\rm max}$ denote the largest comoving volume out of
which an object of absolute magnitude $M$ can be seen, and $V_{\rm min}$ the
analogous if the catalog is also limited at the lower end.
The (true) number of galaxies with absolute magnitude $M$ for a
magnitude limited catalog is:
\begin{equation}
N(M)=\phi(M)[V_{\rm max}(M)-V_{\rm min}(M)],
\label{NM_intr}
\end{equation}
\noindent and the total number of objects with estimated absolute
magnitudes $\cal M$ is:
\begin{eqnarray}
{\cal N}(\cal M) &=& \int {\rm d}M~\phi(M)~\Theta[V_{\rm
                 max}(M),V_{\rm min}(M),M,{\cal M}]  \\
                 &=& \int {\rm d}M~N(M)~\frac{\Theta[V_{\rm
                 max}(M),V_{\rm min}(M),M,{\cal M}]}{[V_{\rm
                 max}(M)-V_{\rm min}(M)]}, \nonumber
\label{NM_obs}
\end{eqnarray}
\noindent where
\begin{equation}
\Theta(V_{\rm max},V_{\rm min},M,{\cal M}) = \int_{D_{\rm L}(V_{\rm
    min})}^{D_{\rm L}(V_{\rm max})} {\rm d}D_{\rm L}~\frac{{\rm
    d}V_{\rm com}}{{\rm d}D_{\rm L}}~p(M-{\cal M}|M,D_{\rm L}).
\label{Theta}
\end{equation}
\noindent Note that since $V_{\rm max}$ and $V_{\rm min}$ are known functions
of $M$, $\Theta$ itself is just a complicated function of $M$ and
${\cal M}$.
\noindent Dividing (\ref{Theta}) by $[V_{\rm max}(M)-V_{\rm min}(M)]$ yields:
\begin{eqnarray}
\frac{\Theta(M,{\cal M})}{[V_{\rm max}(M)-V_{\rm min}(M)]} &=& \int {\rm d}D_{\rm L}~\frac{{\rm
    d}V_{\rm com}/{\rm d}D_{\rm L}}{[V_{\rm max}-V_{\rm
    min}]}~p(M-{\cal M}|M,D_{\rm L}) \nonumber \\
    &=& \int {\rm d}D_{\rm L}~p(D_{\rm L})~p(M-{\cal M}|M,D_{\rm L}) \nonumber \\
    &=& \int {\rm d}D_{\rm L}~p(D_{\rm L})~p({\cal M}|M,D_{\rm L}) \nonumber \\
    &\equiv&  p({\cal M}|M).
\label{pM_obs_given_M_intr}
\end{eqnarray}
\noindent Therefore,  the observed magnitude distribution can be
expressed as a simple one-dimensional deconvolution, namely:
\begin{equation}
{\cal N}({\cal M}) = \int N(M)~p({\cal M}|M)~{\rm d}M.
\label{NM_obs_simpler}
\end{equation}

Along the same lines, use $R$ to denote $\log_{\rm 10}$
of the physical size, and $\cal R$ to denote the estimated size based
on the photometric redshift $\zeta$.
Then it is readily shown that one can also think of ${\cal N}({\cal R})$ as
being a convolution of the true number of objects with size $R$, 
\begin{equation}
{\cal N}({\cal R}) = \int {\rm d}R~N(R)~p({\cal R}|R).
\label{NR_obs_simpler}
\end{equation}
%${\cal R}$ is:
%\begin{eqnarray}
%{\cal N}({\cal R}) &=& \int {\cal N}({\cal M},{\cal
%     R})~{\rm d}{\cal M}~ \nonumber \\
%     &=& \int {\rm d}{\cal M} \int {\rm d}M \int {\rm d}R~N(M,R)~p({\cal M},{\cal R}|M,R) \nonumber \\
%     &=&  \int {\rm d}{\cal M} \int {\rm d}M \int {\rm d}R~N(R)~p(M|R)~p({\cal R}|R)~p({\cal M}|M,R,{\cal R}) \nonumber \\
%     &=&  \int {\rm d}R~N(R)~p({\cal R}|R) \int {\rm d}{\cal M} \int {\rm d}M ~p(M|R)~p({\cal M}|M,R,{\cal R}) \nonumber \\
%     &=& \int {\rm d}R~N(R)~p({\cal R}|R)  \int {\rm d}M~p(M|R)~\int {\rm d}{\cal M} ~p({\cal M}|M,R,{\cal R}) \nonumber \\
%     &\equiv& \int {\rm d}R~N(R)~p({\cal R}|R).
%\label{NR_obs_simpler}
%\end{eqnarray}
%This algebra shows that one can also think of ${\cal N}({\cal R})$ as being
%a convolution of the true number of objects with size $R$.
Direct measurements of the conditional probabilities
allow one to reconstruct the intrinsic distributions from the
observed ones, using a simple one-dimensional deconvolution.
Similarily, a two-dimensional extension of the previous formalism
is necessary if scaling relations are reconstructed from photometric
data (Rossi \& Sheth 2007) \cite{RS2007}.

\begin{figure}[t]
\begin{center}
\psfig{figure=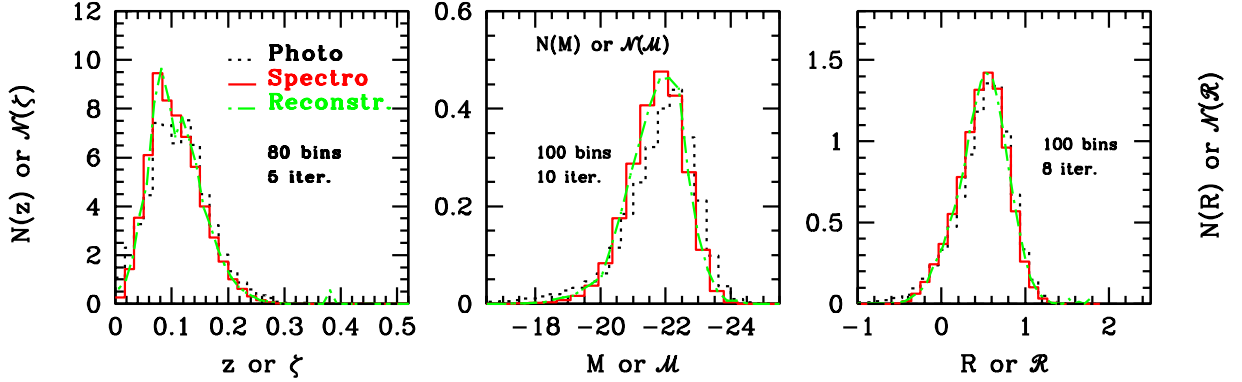,height=2.5in}
\caption{[Left] Observed, intrinsic and reconstructed redshift
          distributions for the SDSS DR6 early-type sample. The dotted histogram was used as a starting guess
          for the one-dimensional deconvolution algorithm. Convergence
         is achieved after a few iterations. [Center] Reconstruction of the intrinsic $N(M)$ distribution
  from the distribution of estimated redshifts. Dotted histogram shows
  the observed absolute magnitude distribution, used as a starting
  guess. Jagged line is the reconstructed intrinsic distribution, after
  $10$ iterations. [Right] Reconstruction of the intrinsic $N(R)$ distribution from the
  distribution of estimated redshifts. Dotted histogram shows the
  observed size distribution, used as a starting
  guess. Jagged lines show the reconstructed intrinsic distribution
  after $8$ iterations. 
\label{fig1}}
\end{center}
\end{figure}

%%%%%%%%%%%%%%%%%%%%%%%%%%%%%%%%%%%%%%%%%%%%%%%%%%%%%%%%%%%%%%%%%%%%%%%%%%%%%%%%%%%%%%%%%%%%%%%%%%%%%%%%%%%%%%%%%%%%%%%%%%

\section{Redshift, magnitude and size distributions. Scaling relations}

Results of applying our deconvolution techniques to the observed
redshift, magnitude and size distributions are shown in Figure \ref{fig1}.
Specifically, the left panel shows the 
photometric or observed redshift distribution (dotted line), the
spectroscopic or intrinsic distribution (solid line) and its
reconstruction after a few iterations (jagged line), based on
the Lucy (1974) \cite{L1974} inversion algorithm. 
The $p(\zeta|z)$ distributions are inferred directly from the SDSS data, and
in our deconvolution code (DeFaST)
we use splines to interpolate for these conditional distributions.
In the same fashion, by measuring the conditional probabilities
$p({\cal M}|M)$ and $p({\cal R}|R)$
directly from the catalog, it is possible to apply
the one-dimensional deconvolution algorithm to reconstruct the magnitude and size
distributions (equations \ref{NM_obs_simpler} and \ref{NR_obs_simpler}).
The central panel shows the 
reconstruction (jagged line) of the intrinsic distribution of
absolute magnitudes (solid histogram) after $10$ iterations. The observed distribution
of $\cal M$ (dotted line) was used as a convenient starting guess in
the deconvolution algorithm.
Similarily, the right panel shows the one-dimensional reconstruction
(jagged line) of the size distribution. The intrinsic distribution of
physical sizes (solid line) is recovered after a few iterations, when the observed distribution
of $\cal R$ (dotted line) is used as a convenient starting guess.

\begin{figure}[t]
\begin{center}
\psfig{figure=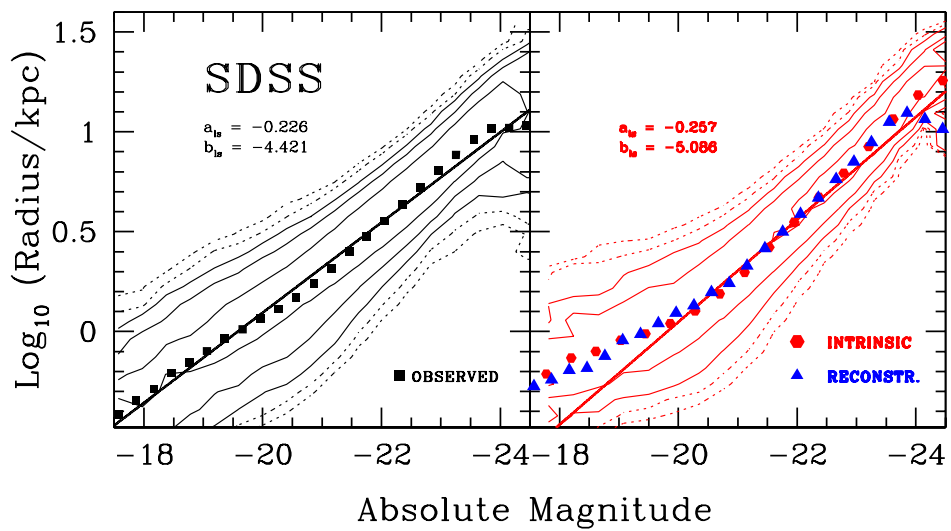,height=3.2in}
\caption{Effect of photo-$z$ on the size-luminosity correlation in
  our SDSS early-type catalog.  In the left panel, contours and solid line
  show the ${\cal R}-{\cal M}$ relation associated with photo-$z$s,
  whereas the right panel shows the intrinsic $R-M$ relation measured
  from spectro-zs.
  Note the bias (shallower slope in panel on left) which
  results from the fact that the photo-$z$ distance error moves
  points down and left or up and right on this plot.
  Squares in left panel show the binned starting guess for the
  2d deconvolution algorithm, triangles in right panel show
  the result after 7 iterations. Circles are the expected binned
  intrinsic relation, obtained from spectroscopic information.
\label{fig2}}
\end{center}
\end{figure}

Although the difference between the intrinsic and observed size distributions
is remarkably small, this departure suffices to bias the
size-luminosity relation -- as presented in
Figure \ref{fig2}.
In fact, photometric redshift errors broaden
both the magnitude and size distributions, but changes to the estimated absolute
magnitudes and sizes are clearly not independent. These correlated
changes have a significant effect on the size-luminosity relation, even
when the brodening of one of the two distributions is not severe.
In our SDSS catalog $\langle {\cal R}|{\cal M} \rangle \propto -0.226$,
whereas $\langle R|M \rangle \propto -0.257$.
In Figure \ref{fig2}
it is shown that the use of photo-$z$
introduces a bias in the size-luminosity relation (shallower slope in panel on left).
Squares in left panel show the binned starting guess for the
two-dimensional deconvolution algorithm, triangles in right panel show
the result after $7$ iterations and circles are the expected binned
intrinsic relation, obtained from spectroscopic
information. Convergence to the true solution is clearly seen.

%%%%%%%%%%%%%%%%%%%%%%%%%%%%%%%%%%%%%%%%%%%%%%%%%%%%%%%%%%%%%%%%%%%%%%%%%%%%%%%%%%%%%%%%%%%%%%%%%%%%%%%%%%%%%%%%%%%%%%%%%%

\section{Summary}

Using a selected sample of early-type galaxies from the SDSS DR6, for which both photo-zs and
spectro-zs are known, we applied our one- and two-dimensional deconvolution
techniques (Sheth 2007 \cite{S2007}; Rossi \& Sheth 2007 \cite{RS2007}) to reconstruct the
unbiased redshift, magnitude and size distributions, as well as
the magnitude-size relation.
We showed that our technique recovers all the true distributions and
the joint relation, to a good degree of accuracy.
We argued that the problem of reconstructing the true magnitude or size distribution
is best thought as a one-dimensional deconvolution problem, and
provided little algebra to show that this is indeed possible.
We showed that even if the distribution of physical sizes is almost unbiased, a bias
in the magnitude distribution sufficies to compromise the
size-luminosity relation in an important way.
We used our 2D technique to correct for this effect.

Although the discussion was phrased mainly in terms of the
luminosity-size relation, the methods developed here are quite general and
can be applied to recover any intrinsic correlations between
distance-dependent quantities (even for $n$-correlated variables).
Potentially, they impact a broader range of studies when at least
one distance-dependent quantity is involved.

%%%%%%%%%%%%%%%%%%%%%%%%%%%%%%%%%%%%%%%%%%%%%%%%%%%%%%%%%%%%%%%%%%%%%%%%%%%%%%%%%%%%%%%%%%%%%%%%%%%%%%%%%%%%%%%%%%%%%%%%%%

\section*{Acknowledgments}

Many thanks to Joey Hyde for a careful reading of the manuscript.

%%%%%%%%%%%%%%%%%%%%%%%%%%%%%%%%%%%%%%%%%%%%%%%%%%%%%%%%%%%%%%%%%%%%%%%%%%%%%%%%%%%%%%%%%%%%%%%%%%%%%%%%%%%%%%%%%%%%%%%%%%


\section*{References}

\begin{thebibliography}{99}

\bibitem{RS2007} Rossi, G., \& Sheth, R.~K.\ 2007, ArXiv e-prints, 710, arXiv:0710.1165
\bibitem{S2007} Sheth, R.~K.\ 2007, \mnras, 378, 709
\bibitem{LCO2007} Lima, M., Cunha, C.~E., Oyaizu, H., Frieman, J., Lin, H., \& Sheldon, E.~S.\ 2008, ArXiv e-prints, 801, arXiv:0801.3822
\bibitem{BSC2000} Budav{\'a}ri, T., Szalay, A.~S., Connolly, A.~J., Csabai, I., \& Dickinson, M.\ 2000, \aj, 120, 1588 
\bibitem{L1974} Lucy L.~B.,\ 1974, \aj, 79, 745

\end{thebibliography}
\end{document}